# Competing contributions of superconducting and insulating states in $Ag_5Pb_2O_6$/CuO composite

**Danijel Djurek**

*Alessandro Volta Applied Ceramics (AVAC),* 10000 Zagreb, Kesten brijeg 5, Croatia
danijel.djurek@zg.t-com.hr

**Abstract:** The composite material, consisting of metallic particles $Ag_5Pb_2O_6$ diluted in insulating CuO matrix, has been investigated in a narrow concentration range separating ballistic tunnel transport from hopping conductivity. The formation of intergrain conducting bridges between metallic particles has been avoided by a careful metallurgic treatment. Resistivity is indicated by temperature dependence $\ln R \sim (T_0/T)^{1/2}$, and this dependence appeals an attention for possible one dimensional conductivity which could be looked upon as a property uniquely associated with an onset of the superconductivity extending up to $T_c \sim 356$ K.

## 1. Introduction

In recent papers [1,2] author emphasized that several dozens of poorly reproducible and unstable superconductivity events [3,4,5] in Ag-Pb-O and Ag-Pb-C-O systems, accompanied with transition temperatures extending up to ambient temperature ($AT_c$), appeared to call into question an interaction between loosely connected grains consisting of a highly conducting ($\sigma \sim$ 20-30 µOhm·cm) Byström-Evers (BE) oxide $Ag_5Pb_2O_6$ [6]. Presumption of intergrain interaction provided important clues for understanding above cited systems, which, in fact, were being BE particles dispersed in insulating CuO, $PbCO_3$ and $Pb_3O_4$ matrices, and such composites seem to be responsible for peculiar physical properties, including confirmation of previously spotted appearances of superconductivity.

In the meantime quality of composite samples $Ag_5Pb_2O_6$/CuO was improved by the use of more advanced metallurgical techniques during the preparation, which in turn increased the stability of recorded data on temperature cycling. This enabled more reliable fits of resistance above $T_c$ to some widely quoted theories which deal with temperature dependent resistance in low dimensional systems, and in this respect samples were prepared with lower concentration of BE oxide exhibiting transition temperature low as possible ($T_c$ = 137 K).

Commonly, an ideal composite structure is supposed as to be nano size spheres of BE oxide dispersed in CuO matrix. In fact, particles are of highly irregular shape which excludes good defined distance between them, and metallurgical procedure has been chosen so as to facilitate the compaction and coagulation of BE particles in CuO matrix to more spherical form.

## 2. Experiment

BE compound $Ag_5Pb_2O_6$ was prepared by mixing in a magnetic stirrer $Ag_2O$ ( Johnson Matthey Co) and $PbO_2$ (Kemika, Zagreb) in stoichiometrc proportion, and ethanol was used as a mixing agent. Subsequent firing was performed at 623 K and $O_2$ pressure 280 bar. $Ag_5Pb_2O_6$ was then reground and mixed in the same way with CuO powder in various weight to weight (w/w) ratios $\gamma = m_{BE}/(m_{BE}+m_{CuO})$. In the first set of preparations grain size of $Ag_5Pb_2O_6$ was 630 nm and that of CuO 315 nm, whereas in next preparations respective sizes were reduced by milling to 65 and 45 nm. The mixture was dried and pelletized in a steel die to the form as shown in the inset of Fig.4. Thickness of pellets was 1,50 mm, and gold wires 120 microns in diameter were used for dc four probe electric resistance measurements performed by the use of Keithley 6221 current source combined with Keithley 2182AE nanovoltmeter. Electric resistance was recorded during heating of the pellet up to 773 K in an evacuated chamber, and annealing at this temperature was performed for 48 hours. At 773 K $Ag_5Pb_2O_6$ decomposed in 10-12 minutes to $5Ag + 2PbO + 2O_2$, which was controlled by the pressure of oxygen released in the chamber. During the course of further annealing in evacuated chamber which lasted at least for 48 hours silver in decomposed BE oxide exhibited a tendency of coagulation to spherical form, which was indicated by an increase of the electric resistance in time, for nearly three orders of magnitude, starting from several tenths ohms recorded just after decomposition. The pellets were then cooled to 583-593 K, and oxygen was introduced in the reaction cell, in order to reestablish the synthesis of BE compound. In these experiments $O_2$ pressure was 1-2 bar, and reaction rate was controlled both, by electric resistance and reverse $O_2$ absorption. Oxygen was then removed by pumping, and sample heated up to 623 K, which increased the range of dc resistance measurements. Above 623 K, even in low oxygen pressures (less than 100 mbar) atmosphere, diffusion between BE particles takes place and conducting bridges are formed, which remove the tunneling gaps, and SC mechanism fails. Furthermore, an optimal $O_2$ pressure must be selected by experience at annealing temperature 583-593 K, in order to prevent formation of such conducting bridges.

Dependence of electric resistance at 375 K on weight to weight ratio $\gamma$ is shown in the inset of Fig. 1. It is evident an abrupt increase of the resistance for nearly three orders of magnitude in a rather narrow concentration range $\Delta\gamma < 0.004$.

The concentration of BE particles in CuO matrix was carefully tuned starting from low values ($\gamma = 0.333$) when samples are insulating at 77 K, up to values when transition to SC was recorded, and $T_c$ was being sufficiently suppressed to leave more extended range of temperature dependent resistance measurements. Temperature dependence of electric resistance for $\gamma = 0.333$ is shown in Fig. 1. Sample is not superconducting down to LN$_2$ temperature, and an attempt to fit the data to temperature dependence of type $\ln R \sim (T_0/T)^\alpha$ failed for any rational value of $\alpha$. For $\gamma = 0.385$ downturn of the electric resistance was observed at T = 137 K (Fig.2), and inset of Fig.2 displays temperature dependence of ac susceptibility. Upon $\ln R$ being fitted to $(T_0/T)^{1/2}$ (Fig.3), two values of $T_0$ may be abstracted; in temperature range spanned between 630 K and room temperature $T_0 = 10^4$ K, whereas fit evaluated by cooling down to 137 K gives $T_0 = 5.3 \cdot 10^4$ K.

Initial powders of BE and CuO were then milled to lower particle sizes. It is noteworthy that concentration at which resistance abruptly increases, as well as the strength of the discontinuous increase are insensitive to the reduction of particle size for an order of magnitude. Fig.4 shows the temperature dependence of the resistance of sample fabricated from milled powders of BE and CuO particles down to the size $a$ = 65 nm and 45 nm respectively ($\gamma = 0.454$). After decomposition of $Ag_5Pb_2O_6$ at 775 K and cooling to 585 K oxygen at 1 bar was introduced in the reaction cell, and reaction back to $Ag_5Pb_2O_6$ was followed by consumption of 0.19 bar $O_2$. Inset represents the specimen arrangement, and voltage contacts were shunted by 2.2 kOhm resistor, in order to remove slight noise. Resistance increases slightly with decreasing temperature down to 351 K and corresponding conductivity at this temperature is $\sigma \sim 370$ Ohm$^{-1}$m$^{-1}$. A rather sharp resistance drop (sampling step was 1 K), firstly observed at 321 K in Ag-Pb-Cu-O mixture [figure 7 in ref 3], is characteristic in these systems. Estimated resistivity at 302 K, for measuring dc current 105 mA, is less than 0.03 μOhm·cm.

### 3. Discussion

The dependence of electric resistance of BE oxide dispersed in CuO matrix on the ratio $\gamma$ indicates that such composite might be considered as an appearance of three possible states. For concentrations $\gamma > 0.50$ percolation regime [7] is dominant, and for $\gamma = 1$ pure polycrystalline $Ag_5Pb_2O_6$ exhibits at room temperature resistivity $\sim$ 20–30 μOhm·cm. In a rather narrow concentration range $0.44 < \gamma < 0.47$ conductivity is governed by tunneling between the grains [8]. Owing to low resistivity in BE particles, being nearly order of magnitude lower than that in bulk polycrystalline samples [9] electronic mean free path is larger than grain diameter $a$, even at ambient temperature, and ballistic tunneling contribution to the conductivity appears to dominate. Conductance at T = 0 is given by Landauer expression $1/R = Te^2/h$ [10,11], and corresponding conductivity is $\sigma = Te^2/ha$, $T$ being tunneling transmission coefficient, and for low intergrain distance (several nanometers) $T \sim 1$. In the experimental configuration presented in the inset of Fig.4, with $a$ = 65 nm, Landauer conductivity at T = 0 should be $\sigma$ = 550 Ohm$^{-1}$m$^{-1}$. This figure nearly matches the value abstracted from Fig.4, supposedly resistance at 356 K is extrapolated down to T = 0, and there might at first sight seem that composite $Ag_5Pb_2O_6$/CuO, in tunneling concentration range, represents an ideal Landauer quantum conductor at T > $T_c$. This regime might be considered upon as being uniquely responsible for an onset of the $AT_c$ superconductivity.

Charge exchange in the ballistic tunneling transport is limited by the minimum energy quantum given by $E_m = e^2/2C$ [12], and $C = 2\pi a \varepsilon_r \varepsilon_0$ is the electric capacity of the supposedly spherical BE particle. At transition temperature Coulomb blockade must be suppressed, which is expressed by the condition $kT_c > E_m$ [13], and required minimum dielectric permeability around the conducting particle should be $\varepsilon_r \sim 0.75$. This rather low estimate results from an overestimated "spherical" value of $a$ evaluated by chemical method dealing with the surface sorption data. Otherwise, the values of dielectric permeability of CuO quoted in the literature extend up to $\varepsilon_r$ = 18.3, and energy stored in the particle is much lower than that required by Coulomb blockade condition.

A considerable increase of the resistance in a rather narrow concentration interval, as it is shown in the inset of Fig. 1, resembles Anderson localization [14,15] in disordered system, when conductivity drops to zero at some critical ratio given by random potential and bandwidth. Conductivity drop is given by $\sigma_{min} \sim 0.16\ e^2/ha$ [16], and assuming $a$ = 65 nm this gives $\sigma_{min}$ = 105 Ohm$^{-1}$m$^{-1}$, which should be compared to 170 Ohm$^{-1}$m$^{-1}$ calculated from the inset of Fig.1. Disagreement comes from the fact that expression for $\sigma_{min}$ includes coordination number z = 6 in denominator which gives pre factor 0.16, and coordination implicitly means 3 dimensional conduction mechanism, being dominant in our case for $\gamma < 0.375$. Conducting particles of BE oxide are randomly distributed, and, beyond the percolation regime, small distance between particles favors randomly distributed potential barriers which might be approximated by delta-functions. In fact, delta potential wells form topological disorder since mean distance between them is nearly constant.

Two types of conductance may appear through delta barriers, quantum tunneling and phonon assisted hopping. The latter mechanism is dominant in the region close to the conductivity drop, and temperature dependence of resistance ln$R$ ~ $(T_0/T)^{1/2}$ sounds for a one dimensional conductivity regime, also reported in some materials having 3 dimensional bulk structure [17]. The lowest possible Coulomb energy stored by conducting BE particle requires an exchange of one electron among adjacent particles, and possible mechanism of one dimensional conductivity is shown in the inset of Fig.3. Gauge invariance requires that one dimensional conducting chains exhibit either localization, that is insulating state at T = 0, or superconductivity. Former case has already been put forward by Thouless [18,19], and both phenomena seem to be in a tight competition, an appearance being in common to some other SC systems when superconductivity is achieved by starting from an insulating state and changing one tuning parameter, like concentration in $La_{2-x}Sr_xCuO_4$.

## 4. Conclusion

A metallurgical treatment of the composite $Ag_5Pb_2O_6$/CuO is described giving rise to the formation of more spherical conducting particles $Ag_5Pb_2O_6$, and separated by an insulating gap of several nanometers in width. Annealing temperature and oxygen pressure were chosen as to prevent the formation of conduction bridges between $Ag_5Pb_2O_6$ particles.

The resistive properties of the composite were investigated in the narrow concentration range which separates the ballistic tunneling transport between conducting particles from the hopping conductivity, the latter being indicated by sharp resistive transition to an insulating state when concentration of conducting particles in the composite is further reduced. For lower concentrations of conducting particles 3 dimensional hopping mechanism dominates , and on the borderline between hopping and tunneling regime system gradually converts to one dimensional conduction mechanism indicated by temperature dependence of electric resistance ln$R$ ~ $(T_0/T)^{1/2}$. This dependence, traced in several independent experiments, is uniquely associated with the onset of superconductivity at ambient temperatures.

**6. Figure captions**

**Fig. 1** Temperature dependence of the electric resistance of $Ag_5Pb_2O_6$/CuO composite for concentration $\gamma$ = 0.333. Inset shows dependence of the electric resistance on $\gamma$, measured at 375 K.

**Fig. 2** Temperature dependence of electric resistance of $Ag_5Pb_2O_6$ composite for $\gamma$ = 0.385. Inset shows temperature dependence of ac magnetic susceptibility.

**Fig. 3** Plot of ln*R* from Fig.2 versus temperature. Inset represents visualization of possible mechanism of one dimensional electric transport, when electronic mean free path exceeds diameter of conducting particles.

**Fig. 4** Temperature dependence of electric resistance of $Ag_5Pb_2O_6$/CuO composite for $\gamma$ = 0.454. Inset shows specimen arrangement in four probe resistance measurement.

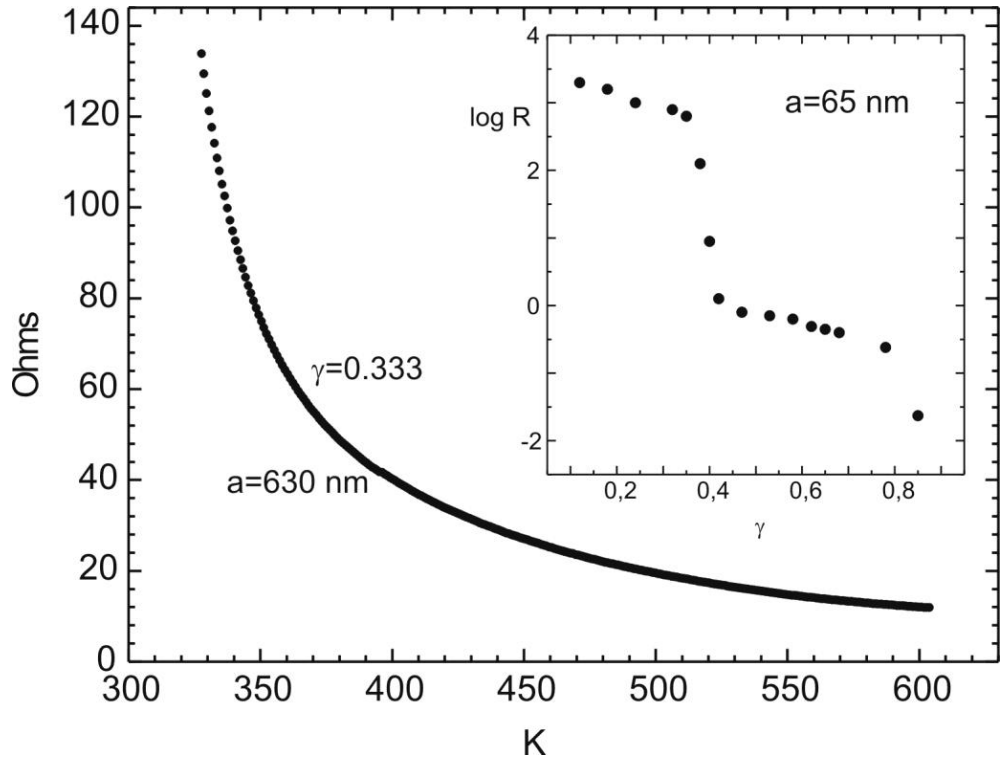

**Figure 1**

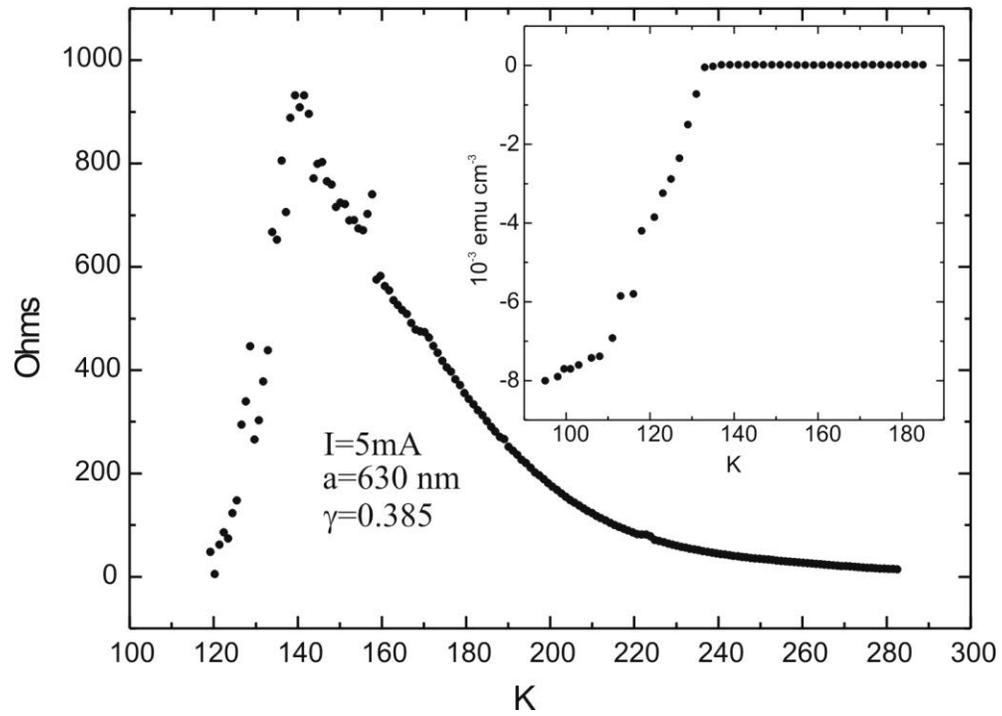

**Figure 2**

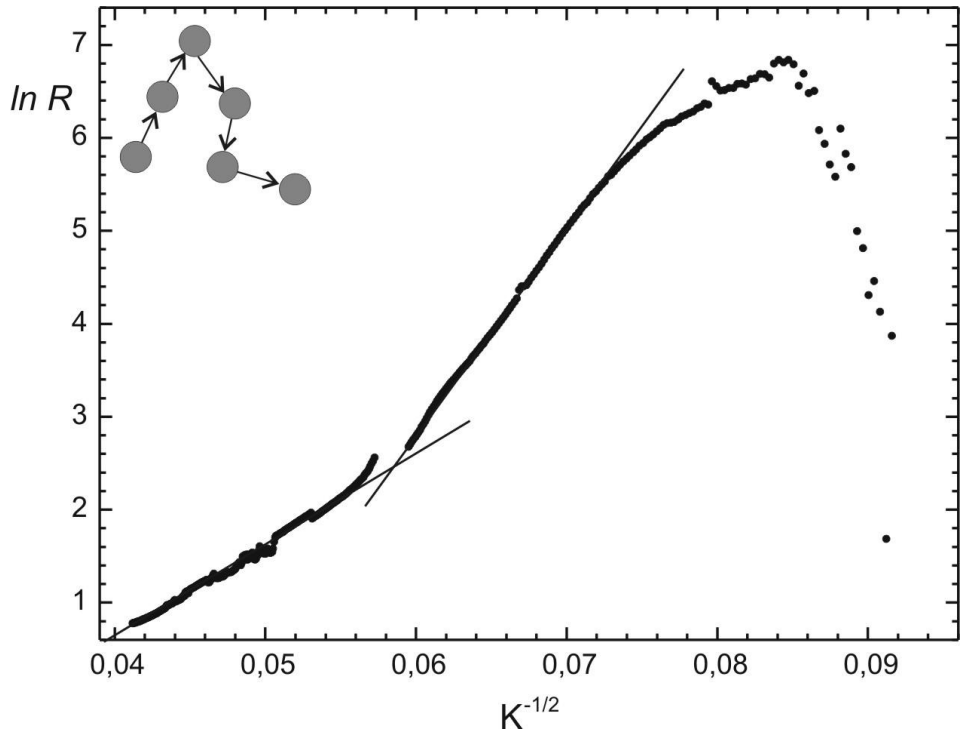

**Figure 3**

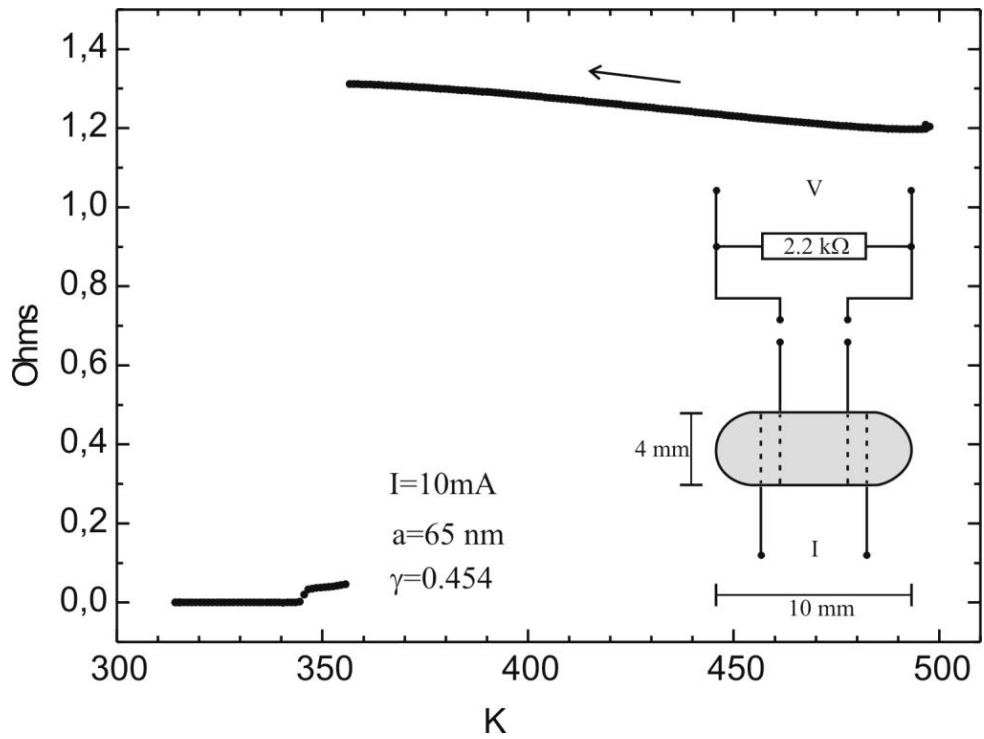

**Figure 4**